\documentstyle[12pt]{article}

\input epsf

\begin{document}
\begin{center}
\Large{{\bf Consequences on variable $\Lambda$-models from distant Type Ia
supernovae and compact radio sources}}
\end{center}

\begin{center}
 R. G. Vishwakarma

\vspace{.2cm}
{\sf IUCAA, Post Bag 4, Ganeshkhind, Pune 411 007, India\\
E-mail: vishwa@iucaa.ernet.in}
\end{center}

\vspace{.7cm}
\noindent
{\bf Abstract.}
We study the magnitude-redshift relation for the Type Ia supernovae data
and the angular size-redshift relation for the updated compact radio
sources data (from Gurvits et al) by considering four variable
$\Lambda$-models: $\Lambda\sim S^{-2}$, $\Lambda\sim H^2$,
$\Lambda\sim \rho$ and $\Lambda\sim t^{-2}$.

It is found that all the variable
$\Lambda$-models, as well as the constant $\Lambda$-Friedmann
model, fit the supernovae data equally well with $\chi^2/\mbox{dof}\approx 1$
and require  non-zero, positive values of $\Lambda$ and an accelerating
expansion of the universe. The estimates of the density parameter for the
variable $\Lambda$-models are found higher than those for the constant
$\Lambda$-Friedmann model.

From the compact radio sources data, it is found, by assuming the
no-evolution hypothesis, that the Gurvits et al' model
(Friedmann model with $\Lambda=0$) is not the best-fitting model
for the constant $\Lambda$ case. The best-fitting Friedmann model (with
constant $\Lambda$) is found to be a low density, vacuum-dominated
accelerating universe. The fits of this data set to the (variable,
as well as, constant $\Lambda$-) models are found very good with
$\chi^2/\mbox{dof}\approx 0.5$ and require  non-zero, positive values
of $\Lambda$ with either sign of the deceleration parameter.
However, for realistic values of the matter density parameter, the only interesting
solutions are (a) estimated from the supernovae data: the best-fit solutions for
 the flat models (including the constant $\Lambda$ case); (b) estimated from the
radio sources data:  
the global best-fit solutions for the models $\Lambda\sim H^2$ and
$\Lambda\sim \rho$, the best-fit solution for the flat model 
with $\Lambda=$ constant
and the Gurvits et al' model.

It is noted that, as in the case of recent CMB analyses, the data sets seem to favour a spherical
universe ($k>0$).

\vspace{.5cm}

\newpage
\noindent
{\bf 1. Introduction}

\noindent
 The cosmic distance measures, for example, the {\it luminosity
distance} and the {\it angular size distance}, depend
sensitively on the spatial curvature and the expansion dynamics
of the models and consequently on the parameters of the models.
For this reason, the magnitude-redshift ($m$-$z$) relation for a
distant {\it standard candle}  and the angular size-redshift
($\Theta$-$z$) relation for a
distant {\it standard measuring rod} have been proposed as potential
tests for cosmological models and play crucial role in determining
cosmological parameters.

The observations on magnitude and redshift of
Type Ia supernovae made, independently, by Perlmutter et al (1999)
and Riess et al (1998) appear to suggest that our universe may be
accelerating with a large fraction of the cosmological density in
the form of the cosmological $\Lambda$-term.
Their findings arise from the study of more than 50
Type Ia supernovae with redshifts in the range $0.16\leq z\leq $ 0.83
and suggest Friedmann models with negative pressure-matter such as a
cosmological constant, domain walls or cosmic strings (Vilenkin 1985,
Garnavich et al 1998).

On the other hand, models with a dynamic cosmological term
$\Lambda (t)$ has been considered in numerous papers to explain
the observed small value of $\Lambda$, which is about
120 orders of magnitude below the value for the vacuum energy density
predicted by quantum field theory (Wienberg 1989, Carroll et al 1992,
Sahni and Starobinsky 2000).
It has been argued that, due to the coupling of the dynamic degree of
freedom with the matter fields of the universe, $\Lambda$ relaxes
to its present small value through the expansion of the universe
and the creation of photons. This approach is essentially phenomenological
in nature but explains, in a natural way, the present small value of
$\Lambda$ which might be large in the early universe. From this point of
view, the cosmological constant is small because the universe is old.

As the dynamics of the variable $\Lambda$-models depends sensitively on the
chosen dynamic law for the variation of $\Lambda$ and, in general,
 becomes altogether
different from the dynamics of the corresponding constant $\Lambda$-models,
there is no reason to believe that the observations of distant objects
would also agree with the  variable $\Lambda$-models, given that they agree
with the corresponding constant $\Lambda$-ones, especially for the same
estimates of the parameters.
In this view, it would be worth while to test the consistency of
these observations with the variable $\Lambda$-models and find the
estimates of different cosmological parameters required by these models.

In this paper, we consider the data on the magnitude and redshift of
Type Ia supernovae from Perlmutter et al and the data on the angular
size and redshift of compact radio sources updated and extended recently
by Gurvits et al (1999) and study the $m$-$z$ and the $\Theta$-$z$ relations
in some decaying $\Lambda$-models. For this purpose, we consider the
following four cases of the phenomenological decay of $\Lambda$:

case (1): $\Lambda \sim S^{-2}$,

case (2): $\Lambda \sim H^2$,

case (3): $\Lambda \sim \rho$,

case (4): $\Lambda \sim t^{-2}$,

\noindent
where $S$, $H$, $\rho$ and $t$ are, respectively, the scale factor of the
R-W metric, the Hubble parameter, the energy density and the cosmic time.
These are among the main dynamical laws one finds in the literature
proposed for the decay of $\Lambda$. Case (1), which has been the subject
of most attention (Ozer and Taha 1987; Chen and Wu 1990;
Abdel-Rahman 1992; Abdussattar and Vishwakarma 1996, 1997; to mention a few),
was originally
proposed by Chen and Wu through dimensional arguments made in the spirit
of quantum cosmology. It is interesting that this ansatz also comes out
as a natural consequence of incorporating the contracted
Ricci-collineation along the fluid flow vector in the Friedmann model
 (Abdussattar and Vishwakarma 1996; Vishwakarma 1999, 2000a).

The case (2) has been proposed from the similar dimensional arguments as
made by Chen and Wu (Carvalho et al 1992, Waga 1993) and also from other
 arguments (Lima and Carvalho 1994) and has also been considered
by several other authors, e.g., Salim and Waga (1993), Arbab and
Abdel-Rahman (1994), Wetterich (1995), Arbab (1997). In view of the
present estimates of $\Lambda$
 being of the order of $H^2_0$, this ansatz seems as a
 natural dynamic law for the decay of $\Lambda$. (The subscript zero
 denotes the value of the quantity at the present epoch).

The dynamical law mentioned in the case (3) has been obtained on the
dimensional ground by suggesting that $\Lambda \sim \Omega H^2$ which
leads to $\Lambda \sim \rho$, (Vishwakarma 2000b), where $\Omega$
is the density of the matter in units of the critical density, i.e.,
$\Omega\equiv \rho/\rho_c =8\pi G\rho/3H^2$ (note that in our notation,
$\Omega$ has contributions from baryons, neutrinos, radiation, etc. and dark
matter {\it but not from $\Lambda$}).

The case (4) has also been considered by several authors by
imposing supplementary conditions which are equivalent to assuming
a power law for the scale factor (Endo and Fukui 1977, Lau 1985,
Bertolami 1986, Berman and Som 1990, Berman 1991, Beesham 1994,
Lopez and Nanopoulos 1996).

\vspace{1cm}
\noindent
{\bf 2.  General Features of the models}

\noindent
In order to find the dynamics of the models resulting from the four
cases of $\Lambda$ mentioned above and then to compare our results
with the earlier works, we consider the homogeneous and isotropic
models in Einstein's theory. We assume that the matter source in the
universe is described by the perfect fluid energy momentum tensor
\begin{equation}
T^{ij}= (\rho +p)u^iu^j + pg^{ij},
\end{equation}
where $\rho$ is the density of baryons, plus neutrinos, plus radiation,
etc., plus dark matter. The Einstein field equations are, then, given by

\begin{equation}
R^{ij}-\frac{1}{2} R g^{ij}=-8\pi G\left[T^{ij}-
\frac{\Lambda (t)}{8\pi G} ~ g^{ij}\right].
\end{equation}
For simplicity we consider units with $c=1$.
In view of the Bianchi identities, this equation suggests that $\Lambda$
would become constant for $T^{ij}=0$. Also, $\Lambda$ must be a constant
if $T^{ij}$ is conserved separately (we consider $G$ as a true constant).
This implies that existence of some non-vanishing
matter, however small, is necessary to kick $\Lambda$ to start
varying. In this connection, we recall that a decaying $\Lambda$ has led
to the possibility of non-singular
models in which initially all the energy is locked up in potential form
in the curvature of spacetime and as the universe evolves, curvature
unfolds and
creates matter (Ozer and Taha 1987, Abdussattar and
Vishwakarma 1996). In this view, this primordial tiny matter (required
by a decaying $\Lambda$) in the otherwise empty baby
universe might serve the purpose of  primordial seeds for structure
formation.

We also observe from equation (2) that the conserved quantity is now
 $[T^{ij}- \{\Lambda (t)/8\pi G\} g^{ij}] \equiv T^{ij}_{{\rm total}}$
and not
$T^{ij}$ and $\Lambda$ separately as would have been the case, had
$\Lambda$ remained a true constant. Thus the incorporation of a variable
$\Lambda$ in the
Einstein field equations is equivalent to postulating an additional source
$-(\Lambda/8\pi G)g^{ij}\equiv T^{ij}_v$ as the energy momentum
tensor of vacuum with the energy density of vacuum
$\rho_v = -p_v =\Lambda/8\pi G$.

For the Robertson-Walker metric
\begin{equation}
ds^2=-dt^2 + S^2(t) \left\{\frac{dr^2}{1-kr^2} +
r^2(d\theta^2+\sin^2\theta d\phi^2)\right\}
\end{equation}
and the usual barotropic equation of state of the matter source

\begin{equation}
p=w\rho, ~ ~ ~ 0\leq w=\mbox{constant}\leq 1,
\end{equation}
equation (2) yields the following two independent equations:
the Friedmann equation

\begin{equation}
\frac{\dot S^2}{S^2}+\frac{k}{S^2}=\frac{8\pi G}{3}\rho+\frac{\Lambda}{3}
\end{equation}
and the Raychaudhuri  equation

\begin{equation}
-\frac{\ddot S}{S}=\frac{4\pi G}{3}(1+3w)\rho-\frac{\Lambda}{3}.
\end{equation}
It is common in cosmology to consider the constant $w=1/3$ for the
radiation-like case and $w=0$ for the dust-like case. Recently
$-1<w<0$ has been suggested by postulating an `x-fluid' (Turner and
White 1997, Zlatev et al 1999) to make up the
long-sought dark matter. However, these new species of matter with
negative pressure are equivalent to a variable cosmological term and
will not be consider here.
The divergence of equation (2) leads to

\begin{equation}
\dot\rho_{{\rm tot}}+3(\rho_{{\rm tot}}+p_{{\rm tot}})\frac{\dot S}{S}=0,
\end{equation}
where $\rho_{{\rm tot}}=\rho+\rho_v$ and $p_{{\rm tot}}=p-\rho_v$.
Equation (7) can also be written as
\begin{equation}
\rho=C S^{-3(1+w)} - \frac{S^{-3(1+w)}}{8\pi G} \int \dot{\Lambda}
S^{3(1+w)}dt, ~ ~ C=\mbox{constant}.
\end{equation}
This indicates that a decaying $\Lambda$ $(\dot{\Lambda}<0)$ makes a
positive contribution to $\rho$. Equations (5) and (6) can be used to
obtain

\begin{equation}
2\frac{\ddot S}{S} + (1+3w)\left(\frac{\dot S^2}{S^2}+\frac{k}{S^2}
\right)- (1+w)\Lambda=0,
\end{equation}
which gives the dynamics of the scale factor for a particular
$\Lambda$.

We shall now describe, briefly, the models resulting from the different
dynamical laws for the decay of $\Lambda$ mentioned in the previous
section.

\newpage
\vspace{.8cm}
\noindent
{\bf Case (1):}

\noindent
We consider

\begin{equation}
\Lambda=\frac{n_1}{  S^2},
\end{equation}
where the constant $n_1$ is a new cosmological parameter replacing the
original $\Lambda$ and is to be determined from the observations.
Equations (8), (9) and (10) give the dynamics of the model as

\begin{equation}
\rho=C_1 S^{-3(1+w)} + \frac{n_1S^{-2}}{4\pi G(1+3w)},
 ~ ~ C_1=\mbox{constant}
\end{equation}
and

\begin{equation}
\dot{S}^2=\frac{8\pi G C_1}{3} ~ S^{-(1+3w)} + \frac{n_1(3+w)}{3(1+w)}
-k.
\end{equation}

\vspace{.8cm}

\noindent
{\bf Case (2):}

\noindent
We now consider

\begin{equation}
\Lambda=n_2 ~ H^2,
\end{equation}
where, as in the previous case, the parameter $n_2$ is to be determined
from the observations.
The use of (13) in equations (5) and (8), leads to the
following equations

\begin{equation}
\rho=C_2 S^{(n_2-3)(1+w)} - \frac{n_2 k S^{-2}}{4\pi G[(3-n_2)(1+w)-2]}, ~
~ C_2=\mbox{constant}
\end{equation}
and
\begin{equation}
\dot{S}^2=\frac{8\pi GC_2}{(3-n_2)} S^{(n_2-3)(1+w)+2} -
\frac{(1+3w)k}{[(3-n_2)(1+w)-2]},
\end{equation}
which determine the dynamics of the model.

\vspace{.8cm}
\noindent
{\bf Case (3):}

\noindent
We now consider the case

\begin{equation}
 \Lambda=n_3 \rho,
\end{equation}
where the parameter $n_3$, describing the model,
is to be determined from the observations as in the earlier cases (1)
and (2).

\noindent
The dynamics of the models can be obtained from equations (5), (8) and
(16) as

\begin{equation}
\rho = C_3 S^{-3(1+w)/(1+n)}, ~ ~ C_3=\mbox{constant}\geq 0
\end{equation}
and
\begin{equation}
\dot{S}^2=\frac{8\pi GC_3}{3}(1+n) ~ S^{(2n-1-3w)/(1+n)} - k,
\end{equation}
where $n\equiv n_3/8\pi G$.

\vspace{.8cm}
\noindent
{\bf Case (4):}

\noindent
Finally we consider the case

\begin{equation}
 \Lambda=\frac{n_4}{ t^2},
\end{equation}
where the parameter $n_4$ describes this family of models.
By using (19), equation (9) can be written as

\begin{equation}
\frac{\ddot x}{x}=\frac{3(1+w)^2 n_4}{4t^2},
\end{equation}
where the new variable $x$ is related to $S$ by means of the
transformation

 \begin{equation}
 \frac{\dot S}{S}=\frac{2}{3(1+w)}\frac{\dot x}{x}
 \end{equation}
implying that $S\sim x^{2/3(1+w)}$. Here we have restricted ourselves to
the flat model only as has been done by the other authors. By assuming
$t=e^y$, equation (20) can be further transformed to

 \begin{equation}
 \frac{\mbox{d}^2 x}{\mbox{d} y^2}-\frac{\mbox{d}x}{\mbox{d} y}
 -\frac{3(1+w)^2n_4}{4}x=0,
 \end{equation}
 which gives a physically viable solution as

\begin{equation}
S \sim t^{n_5}, ~ ~ \mbox{with} ~ ~ n_5=\frac{1+\sqrt{1+3(1+w)^2n_4}}{3(1+w)},
\end{equation}
provided $n_4>-1/3(1+w)^2$. Other solutions (for $n_4\leq-1/3(1+w)^2$)
are not interesting on the ground of age considerations (Overduin
and Cooperstock 1998).  Solution (23) is consistent with the different
power-law solutions obtained by various authors by assuming different
supplementary conditions, as we have mentioned in the previous
section. However, in view of $H\sim t^{-1}$ as is implied by equation
(23), this case reduces to the case (2) of $\Lambda \sim H^2$ and will
not be considered separately further.

While comparing the models with observations, we shall be mostly interested
in the effects which occurred at redshift $z<5$. We, therefore, neglect
radiation and consider $w=0$ in the following. Equations (5) and (6), then,
can be recast in the following forms to give the relative contributions of
different cosmological parameters at the present epoch:

\begin{equation}
1+\Omega_{k0} = \Omega_{0} + \Omega_{\Lambda 0},
\end{equation}

\begin{equation}
2[q_0 + \Omega_{\Lambda 0}]=\Omega_{0},
\end{equation}
where $\Omega_\Lambda \equiv \Lambda/3H^2$ and $\Omega_k\equiv k/S^2 H^2$
are, respectively, the energy density of vacuum and that associated
with curvature in units of the critical density and  $q$ is the deceleration
parameter.

It would be worth while to find out if there exist some characteristics of
the models, other than the different dependences of $\Lambda$, which
clearly distinguish the evolution of the models discussed above. The
dynamics of the models, described by equations (12), (15) and (18),
depend on the parameters $n_i$ and $C_i$. In order to make the comparison
possible, we write them in terms of $H_0$, $\Omega_{0}$ and
$\Omega_{\Lambda 0}$ as follows:

\begin{equation}
n_1=3H_0^2 S_0^2 \Omega_{\Lambda 0}, ~ ~ ~ C_1=\frac{3H_0^2}{8\pi G}
[\Omega_0 - 2 \Omega_{\Lambda 0}] S_0^3;
\end{equation}

\begin{equation}
n_2=3 \Omega_{\Lambda 0}, ~ ~ ~ C_2=\frac{3H_0^2}{8\pi G}
\left[\frac{(1-\Omega_{\Lambda 0})(\Omega_0 - 2 \Omega_{\Lambda 0})}{
1-3\Omega_{\Lambda 0}}\right] S_0^{3(1-\Omega_{\Lambda 0})};
\end{equation}

\begin{equation}
n_3=8\pi G\frac{\Omega_{\Lambda 0}}{\Omega_0}, ~ ~ ~ C_3=\frac{3H_0^2}{8\pi G}
\Omega_0 S_0^{3\Omega_0/(\Omega_0 +\Omega_{\Lambda 0})},
\end{equation}
where $S_0=H_0^{-1}\sqrt{k/\Omega_{k0}}$ with $\Omega_{k0}$ given by
equation (24). Equations (12), (15) and (18) then reduce, respectively, to the
following.

\noindent
$\Lambda\sim S^{-2}$:
\begin{equation}
\dot S^2=H_0^2S_0^2\left[
(\Omega_0 - 2 \Omega_{\Lambda 0})\frac{S_0}{S} + (1-\Omega_0 +
 2 \Omega_{\Lambda 0})\right];
\end{equation}

\noindent
$\Lambda\sim H^2$:
\begin{equation}
\dot S^2=H_0^2S_0^2\left[\frac{
\Omega_0 - 2 \Omega_{\Lambda 0}}{1-3\Omega_{\Lambda 0}}\left(
\frac{S_0}{S}\right)^{1-3\Omega_{\Lambda 0}} +\frac{
1-\Omega_0 - \Omega_{\Lambda 0}}{1-3\Omega_{\Lambda 0}}\right];
\end{equation}

\noindent
$\Lambda\sim \rho$ :
\begin{equation}
\dot S^2=H_0^2S_0^2\left[(\Omega_0
+\Omega_{\Lambda 0})\left(\frac{S_0}{S}\right)^{(\Omega_0 -
2 \Omega_{\Lambda 0})/(\Omega_0+\Omega_{\Lambda 0})} +
1-\Omega_0 - \Omega_{\Lambda 0}\right].
\end{equation}
It is now clear from equations (29)-(31) that the dynamics of the models
depend only on the values of $\Omega_0$ and $\Omega_{\Lambda 0}$.
Also the relative behaviour of $S$ from these equations is quite arbitrary
and changes with different pairs of $\Omega_0$ and $\Omega_{\Lambda 0}$.
It thus seems that in general there is no feature which distinguishes
the models. We also note that for $\Omega_0=2\Omega_{\Lambda 0}$, the
models become identical reducing to
$S\sim t$ and $\rho\sim\Lambda\sim t^{-2}$. Also, for the trivial case
$\Lambda=0$, the models become identical reducing, obviously, to the
standard FRW models with zero $\Lambda$.

We shall now derive the $m$-$z$ and $\Theta$-$z$ relations in
the models in order to compare them with observations, which will be done
in the following sections.

\vspace{1cm}
\noindent
{\bf 3.  Study of magnitude-redshift relation in the models}

\noindent
Suppose that the observer at $r=0$ and $t=t_0$ receives the light
emitted at $t=t_1$ from a source of absolute luminosity $L$
located at a radial distance $r_1$. The cosmological redshift $z$
of the source is related with $t_1$ and $t_0$
by $1+z=S(t_0)/S(t_1)$. If the (apparent) luminosity of the source
measured by the observer is $l$,
the {\it luminosity distance} $d _{\mbox{{\scriptsize L}}}$ of the
source, defined by

\begin{equation}
l \equiv \frac{L}{4\pi d _{\mbox{{\scriptsize L}}}^2},
\end{equation}
is then given by

\begin{equation}
d _{\mbox{{\scriptsize L}}}=(1+z) S_0 ~ r_1.
\end{equation}
For historical reasons, the absolute and apparent
luminosities  $L$ and $l$ are defined, respectively,  in terms of
the absolute and apparent magnitudes $M$ and $m$ as
$L=10^{-2M/5}\times 3.02 \times 10^{35}$ erg s$^{-1}$ and $l=10^{-2m/5}
\times 2.52 \times 10^{-5}$ erg cm$^{-2}$ s$^{-1}$ (Weinberg 1972). When
written in terms of $M$ and $m$, equation (32) yields

\begin{equation}
m(z;{\cal M}, \Omega_0,\Omega_{\Lambda 0})={\cal M} +
5 \mbox{log}_{10}\left[\frac{{\cal D} _
{\mbox{{\scriptsize L}}}(z; \Omega_0,\Omega_{\Lambda 0})}
{1 \mbox{Mpc}}\right],
\end{equation}
where ${\cal M}\equiv M-5\mbox{log}_{10}H_0 +25$ and ${\cal D}_{
\mbox{{\scriptsize L}}}(z; \Omega_0,\Omega_{\Lambda 0})
\equiv H_0 ~ d _{\mbox{{\scriptsize L}}}(z; \Omega_0,
\Omega_{\Lambda 0}, H_0)$
is the dimensionless {\it luminosity distance}. By using equation
(3), the coordinate distance $r_1$, appearing in equation (33),
yields

\begin{equation}
\psi (r_1) =\int_{S_0/(1+z)}^{S_0} \frac{dS}{S\dot S}
\end{equation}
with

\[
\psi(r_1)=\sin^{-1} r_1, ~ ~ ~ ~ k=1
\]
\[~ ~ ~ ~  ~ ~ ~ =r_1, ~ ~ ~  ~ ~  ~ ~ ~ ~ ~ ~ ~ k=0
\]
\begin{equation}
~ ~ ~ ~ ~ ~ ~ ~ ~ ~ ~ =\sinh^{-1} r_1, ~ ~ ~ k=-1.
\end{equation}

The {\it luminosity distances} for the different models, described
for the cases (1), (2) and (3), will be
given by calculating the right hand side of equation (35) for these
models. By using equations (29)-(31), equation (35) can
be written, for the three models, as

\begin{equation}
\psi(r_1) = \frac{1}{S_0 H_0} I_j,
\end{equation}
where $I_j$ ($j=1$, 2 and 3 correspond to, respectively, the
three models described for the cases (1), (2) and (3)) is given by

\begin{equation}
I_1= \int_0^z [(2\Omega_{\Lambda0}-\Omega_0 + 1)(1+z')^2
-(2\Omega_{\Lambda0}-\Omega_0)(1+z')^3 ]^{-1/2} dz',
\end{equation}

\begin{equation}
I_2=\int_0^z \left[
\frac{(\Omega_0-2\Omega_{\Lambda0})}{(1-3\Omega_{\Lambda0})} ~
(1+z')^{3(1-\Omega_{\Lambda0})} +
\frac{(1-\Omega_0-\Omega_{\Lambda0})}{(1-3\Omega_{\Lambda0})} ~
(1+z')^2\right]^{-1/2}dz'
\end{equation}
and

\begin{equation}
I_3=\int_0^z
[(\Omega_0+\Omega_{\Lambda0})(1+z')^{3\Omega_0/(\Omega_0+\Omega_{\Lambda0})}
+(1-\Omega_0-\Omega_{\Lambda0})(1+z')^2 ]^{-1/2} dz'.
\end{equation}
(For convenience, we also write the similar integral $I_c$ for the
constant $\Lambda$-Friedmann models obtained as
\[
I_c=\int_0^z[(1+\Omega_0 z')(1+z')^2 - z'(2+z')\Omega_{\Lambda0}]^{-1/2} dz',
\]
which will be used to compare our results with those for the
constant $\Lambda$-models.)

Equations (33), (36) and (37) can also be combined into a single
compact equation to give

\begin{equation}
{\cal D} _{\mbox{{\scriptsize L}}}(z; \Omega_0,\Omega_{\Lambda 0})=
\frac{(1+z)}{\sqrt{{\cal K}}} ~ \xi\left(\sqrt{{\cal K}} I_j \right),
\end{equation}
where

$\xi(x)=\sin (x)$ with ${\cal K}=\Omega_{k0}$ when $\Omega_{k0}>0$,

$\xi(x)=\sinh (x)$  with ${\cal K} = - \Omega_{k0}$ when
$\Omega_{k0}<0$ and

$\xi(x)=x$ with ${\cal K}=1$ when $\Omega_{k0}=0$.

Thus for given ${\cal M}$, $\Omega_0$ and $\Omega_{\Lambda0}$, equations
(34) and (41) give the predicted value of $m(z)$ at a given $z$. Using
the observed values of the effective magnitude
$m_i^{\mbox{{\scriptsize eff}}}$ (corrected for the width-luminosity
relation) and the same standard errors $\sigma_{z,i}$ and $\sigma_{
m_i^{\mbox{{\scriptsize eff}}}}$ of the $i$th supernova with redshift
$z_i$ as used by Perlmutter et al, we compute $\chi^2$ according to

\begin{equation}
\chi^2=\sum_{i=1}^{54} \frac{[m_i^{\mbox{{\scriptsize eff}}} - m(z_i)]^2}
{(\sigma_{z,i}^2 + \sigma_{m_i^{\mbox{{\scriptsize eff}}}}^2)}.
\end{equation}
The best fit parameters are obtained by minimizing this equation.

We consider the data set on $m^{\mbox{{\scriptsize eff}}}$ and $z$ of 54
supernovae as used by Perlmutter et al in their primary fit C.
This set comprises 38 high-redshift supernovae from the Supernova
Cosmology Project (excluding 2 outliers and 2 likely reddened ones from the
full sample of 42 supernovae) together with 16 low-redshift supernovae from
the Calan-Tololo sample (excluding 2 outliers from the full sample of 18
supernovae). We fit the low- and high-redshift supernovae simultaneously
to equation (34) to estimate the parameters ${\cal M}$, $\Omega_0$
and $\Omega_{\Lambda0}$. This fitting procedure is followed by Perlmutter
et al in their primary fit C. We have listed the best-fit values in Table 1.
The global best-fit parameters have been calculated by giving free rein
to $\Omega_0$, $\Omega_{\Lambda0}$ and ${\cal M}$, i.e., at $54-3=51$
degrees of freedom (dof) whereas the best-fit parameters for flat models are
calculated at 52 degrees of freedom with the constraint
$\Omega_0+\Omega_{\Lambda0}=1$.
For comparison, we have shown the best-fit parameters for the constant
$\Lambda$-models.
The $\chi^2$ values obtained at different
degrees of freedom have also been converted into confidence levels (cl)
[the goodness of fit probability $Q=1-$cl$\%$].
In Figure 1, we have shown the fit of the data to the best-fitting
flat models and compared it with the fit of the data to the
Einstein-de Sitter model.

It may be mentioned that Perlmutter et al have also fitted the data for
only 3 parameters ${\cal M}$, $\Omega_0$ and $\Omega_{\Lambda0}$ and
not for the 4 parameters  ${\cal M}$, $\Omega_0$,
$\Omega_{\Lambda0}$ and $\alpha$ (the slope of the width-luminosity
relation) as mentioned in their paper (from
 a personal discussion with Professor R. S. Ellis, one of the authors
 of the paper). A self-consistent 4-parameter fit has been done by
 Efstathiou et al (1999).

\vspace{.5cm}
\noindent
{\bf Table 1.} Best-fit values of different parameters for three
variable $\Lambda$-models estimated from the Type Ia supernovae data.
For comparison, the best-fit parameters for the constant
$\Lambda$-Friedmann model (Perlmutter et al') and Einstein-de Sitter
model have also been shown.
\begin{center}
\begin{tabular}{lll}
\hline\hline
\multicolumn{1}{c}{Models}&
\multicolumn{1}{c}{Best-Fit Parameters (Flat models)}&
\multicolumn{1}{c}{Global Best-Fit Parameters}\\
 &\begin{tabular}{lllll} $\Omega_0$  & ~ ~ ${\cal M}$ & ~ ~ $\chi^2$
 &  dof & cl($\%$)\end{tabular}&
\begin{tabular}{llllll}$\Omega_0$  & ~ $\Omega_{\Lambda0}$ &
~ ${\cal M}$  & ~ ~ $\chi^2$ & ~dof & cl($\%$)\end{tabular}\\
\hline

$\Lambda \sim S^{-2}$ & \begin{tabular}{lllll}0.49  & 23.97  & 58.97
&52 & ~76 \end{tabular}&
\begin{tabular}{llllll}1.86 & 1.52 & 23.95 & 57.43 &51 &~75 \end{tabular}\\

$\Lambda \sim H^2$ & \begin{tabular}{lllll}0.4 ~& 23.96  & 58.33
&52 & ~75 \end{tabular}&
\begin{tabular}{llllll}0.98 & 1.53  &  23.91 & 56.86 &51 &~73\end{tabular}\\

$\Lambda \sim \rho$& \begin{tabular}{lllll}0.4 ~& 23.96  &  58.33
&52 & ~75\end{tabular}&
\begin{tabular}{llllll}1.62 & 1.59 & 23.93 & 56.91 &51 & ~74\end{tabular}\\

$\Lambda$$=$const& \begin{tabular}{lllll}0.28 & 23.94  &  57.68
&52 & ~73\end{tabular}&
\begin{tabular}{llllll}0.79  & 1.41 & 23.91 & 56.82 &51 &~73\end{tabular}\\

\begin{tabular}{l}$\Lambda=0$,\\$\Omega=1$ \end{tabular}
& \begin{tabular}{lllll} ~  ~ ~ & 24.21  & 92.89
&53 & 99.9\end{tabular}&
\begin{tabular}{llllll} ~ & ~  ~  & ~  ~ & ~& &  \end{tabular}\\
\hline
\end{tabular}
\end{center}

\centerline{{\epsfxsize=14cm {\epsfbox[50 250 550 550]{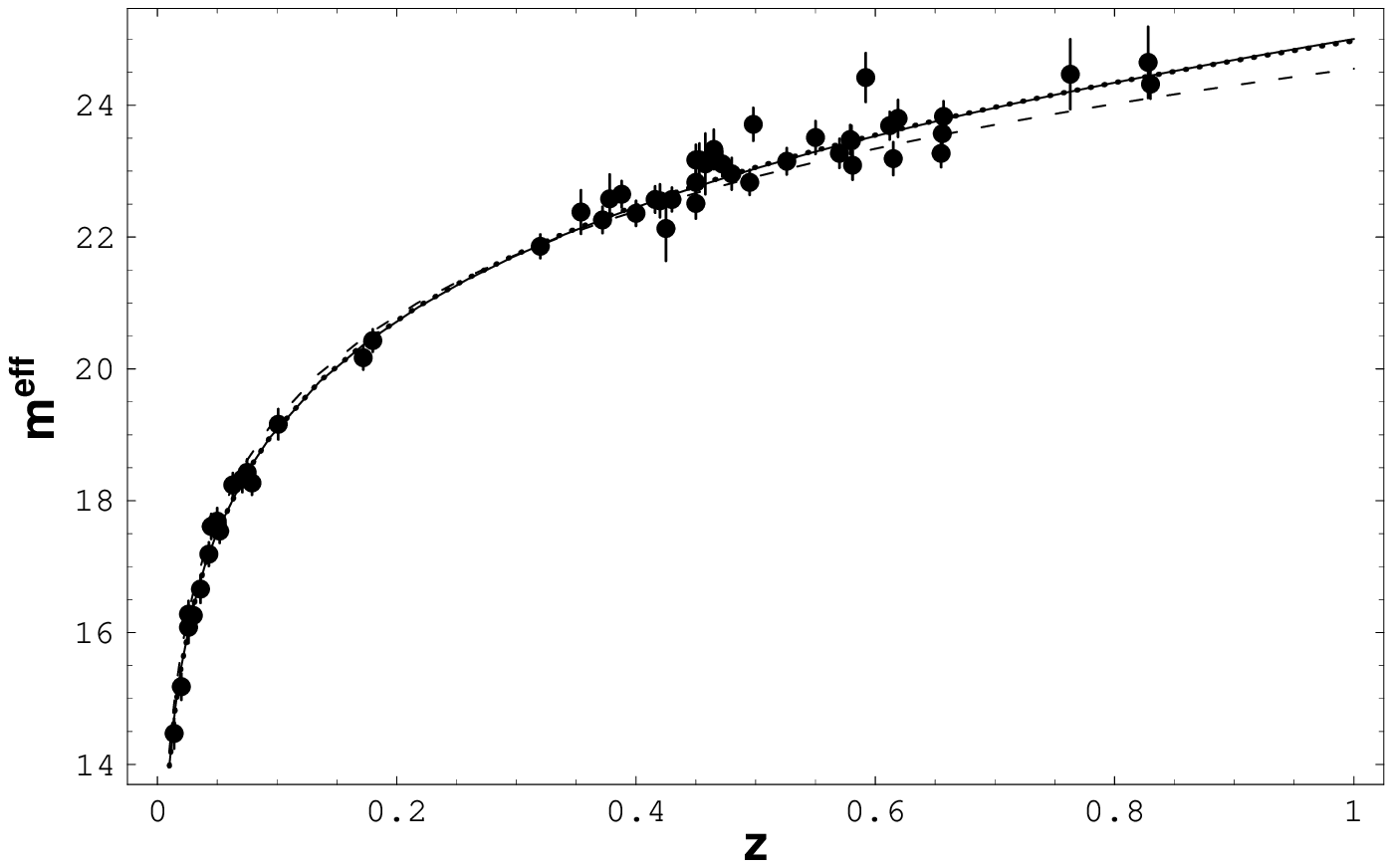}}}}

\noindent
{\bf Figure 1.}
Hubble diagrams for 38 high-redshift and 16 low-redshift supernovae are shown.
The solid curve represents the best-fitting flat model $\Omega_0=0.49$
(for $\Lambda \sim S^{-2}$) and the dotted curve represents the best-fitting
flat model $\Omega_0=0.4$ (for $\Lambda \sim H^2$).
The curves overlap on each other because of the same goodness of fit.
For comparison, the canonical Einstein-de Sitter model ($\Omega_{k}=0$,
$\Omega_{\Lambda}=0$) has also been plotted (dashed curve).

\vspace{.8cm}
 The fits of the data to all the three  variable
$\Lambda$-models are almost as good as to the constant
$\Lambda$-Friedmann model (Perlmutter et al') with $\chi^2/$dof $\approx 1$
(a very good fit indeed) and require non-zero,
positive values of $\Lambda$. It is clear from Table 1 and Figure 1
that the data
do not fit the Einstein-de Sitter model ($\Omega=1, \Lambda=0$).
We note that the
global best-fit values for $\Omega_0$ and $\Omega_{\Lambda0}$ are always
larger than those for the corresponding flat models. Also the estimates
of $\Omega_0$
for the variable $\Lambda$-models are always higher than that for
the constant $\Lambda$-Friedmann model (Perlmutter et al') and hence
require dark matter, though not as much as in the non-flat models.
It is also clear from the Table that for all
the three variable $\Lambda$-models, $\Omega_0/2$ is less than the
corresponding
$\Omega_{\Lambda0}$ (as is the case for the constant $\Lambda$-Perlmutter
et al' model) leading to $q_0<0$.

One may notice that the estimates of the parameters for the flat models
for the cases $\Lambda \sim H^2$ and $\Lambda \sim\rho$ are exactly same.
This simply reveals the fact that for $k=0$, the two models become identical.
This can be checked from equations (17) and (18) which, for $k=0$, give
$S\sim t^{2(1+n)/3(1+w)}$ leading to $\Lambda\sim\rho\sim t^{-2}\sim H^2$.
This may be considered as a cross check for our fitting procedure.

Elsewhere (Vishwakarma 2000a) we have used this data set with a different
fitting procedure for the model $\Lambda \sim S^{-2}$. Instead of fitting
the high and low-redshift data simultaneously, we used low-redshift
measurements to determine ${\cal M}$ and by using this value in equation
(34) we performed a two-parameter fit of the high-redshift supernovae to
estimate $\Omega_0$ and $\Omega_{\Lambda0}$. The best-fit values so obtained
are in good agreement with the estimates shown in Table 1 for that model.

We would like to mention that the supernovae data have also been used in many
studies to place constraints on models with `x-fluid' ($p_x=w\rho_x$ with
$-1< w <0$) or `quintessence' (a slowly varying scalar field) (Waga and Miceli
1998; Efstathiou 1999; Perlmutter, Turner and White 1999;
Podariu and Ratra 1999; and
the references therein). Though `x-fluid' and `quintessence' are formally
equivalent to a variable $\Lambda$, however, these models are essentially
different from our kinamatical `usual' $\Lambda$-models considered in
this paper in the sense that in our models matter is created as a result of
the decaying $\Lambda$, as has been mentioned in section 2, whereas in the
exotic fluid description, the exotic component is conserved. In the present
approach (or in the case of `x-fluid' with $w=-1$), if the $\Lambda$-source
[$\{\Lambda(t)/8\pi G\}g_{ij}$] is conserved separately, then
$\Lambda$ would become constant, as has been mentioned earlier.

\vspace{1cm}
\noindent
{\bf 4.  Study of angular size-redshift relation in the models}

\noindent
The compact radio sources, of angular size of the order of a few
milliarcseconds, which are observed with very long baseline
interferometry (VLBI), are deeply embedded in
the galactic nuclei and  are expected to be free from the
evolutionary effects. These sources are attractive for the
test of cosmological models because they are usually identified with
quasars which, in general, have high redshifts, so the differences
between models may be more easily distinguished than for the extended
double-lobed sources.

Kellermann (1993) used a sample of 79 such
sources and showed that a credible $\Theta$-$z$ relation can emerge.
His main motivation was, however, limited to showing that the resulting
$\Theta$-$z$ diagram was cosmologically credible and compatible with the
Einstein-de Sitter canonical
model with $\Lambda=0$ rather than to be specific about the
acceptable ranges for the various cosmological parameters. Extending his
work, by introducing $\Lambda$ into the picture, Jackson
and Dodgson (1996) showed that the Kellermann's data could also fit very
well with a low density, highly decelerating model having a large negative
cosmological constant.
Later on, a more extensive exercise was carried out by Jackson and Dodgson
(1997) by considering a bigger sample of 256 ultracompact sources
 selected from the compilation from Gurvits (1994). Then,
they concluded that the canonical cold dark matter model with
$\Lambda=0$ was ruled out by the observed relationship at 98.5 percent
level of confidence and low density FLRW models with either sign of
$\Lambda$ were favoured. Elsewhere we (Vishwakarma 2000b) used
the data set from Jackson and Dodgson (1997) in the model resulting
from the case (3) and found the model in good agreement with the data.

Recently this data set has been updated and extended by Gurvits et al
(1999) as a sample of 330 sources (though at the expense of homogeneity)
distributed over a wide range of redshifts $0.011\leq z \leq 4.72$.
In order to minimize any possible
dependence of linear size on luminosity and that of angular size on
spectral index, they discarded lower values of luminosities and extreme
values of spectral indices and selected only 145 sources with total
radio luminosity $Lh^2\geq 10^{26}$ W/Hz and $-0.38\leq$ spectral index
$\leq 0.18$, which are hoped to be free from the evolutionary effects
and hence conceivably comprise a set of standard objects. This sub-sample
was distributed into 12 bins, each bin containing 12-13 sources.
 They studied the $\Theta$-$z$ relation in Friedmann
models with $\Lambda=0$ for this sample and found the best-fit
model as $q_0=0.21$ with no-evolution assumption. However, as we shall
see, the models with a non-zero $\Lambda$ (constant) are also in good
agreement (even better) with the data.
We shall also use this data set to study the  $\Theta$-$z$ relation in
the variable-$\Lambda$ models considered in this paper.
For this purpose, we now derive the $\Theta$-$z$ relation in the models
by assuming that there are no evolutionary effects due to a linear
size-luminosity or linear size-redshift dependence.

The {\it angular diameter distance} $d_{{\rm A}}$ of a light source of
linear size $d$ and angular size $\Theta$, defined by

\begin{equation}
d_{{\rm A}}=\frac{d}{\Theta},
\end{equation}
is related to the {\it luminosity distance} $d_{{\rm L}}$ of the source
by

\begin{equation}
d_{{\rm A}} =\frac{d_{{\rm L}}}{(1+z)^2}.
\end{equation}
By comparing this with equation (33) and using (36), (37) and (43),
we obtain the $\Theta$-$z$ relation in the models as

\begin{equation}
\Theta (z; dh,\Omega_0, \Omega_{\Lambda0})=
\left(100 ~ \mbox{km s}^{-1} \mbox{Mpc}^{-1}\right)
dh \sqrt{{\cal K}} ~ (1+z) ~ / ~ \xi (\sqrt{{\cal K}} ~ I_j),
\end{equation}
where $h$ is the present value of the Hubble parameter in units of
100 km s$^{-1}$ Mpc$^{-1}$.

It is thus clear that for given $dh$, $\Omega_0$ and
$\Omega_{\Lambda0}$, the predicted value $\Theta(z)$, at a given
$z$, is completely determined. Using median values of the grouped
data into 12 bins, as used by Gurvits et al, we calculate the
theoretical $\Theta(z)$ for a wide range of parameters $dh$,
$\Omega_0$ and $\Omega_{\Lambda0}$ and compute  $\chi^2$ according to

\begin{equation}
\chi^2=\sum_{i=1}^{12} \left[\frac{\Theta_i -
\Theta(z_i)}{\sigma_i}\right]^2,
\end{equation}
which measures the agreement between the theoretical $\Theta(z)$ and the
observed value $\Theta_i$ with the  errors $\sigma_i$
of the $i$th bin in the sample. The maximum
likelihood estimates of the parameters are obtained by minimizing
$\chi^2$ as usual.
We have listed our results in Table 2 below.

\newpage
\vspace{.5cm}
\noindent
{\bf Table 2.} Best-fit parameters for the Friedmann model (with
constant $\Lambda$) and three variable $\Lambda$-models estimated
from Gurvits et al' data. For comparison, the best-fit parameters
for Gurvits et al' model have also been shown. (The values of $dh$ are
shown in unit of 1 pc.)

\begin{center}
\begin{tabular}{lll}
\hline\hline
\multicolumn{1}{c}{Models}&
\multicolumn{1}{c}{Best-Fit Parameters (Flat models)}&
\multicolumn{1}{c}{Global Best-Fit Parameters}\\
 &\begin{tabular}{lllll} $\Omega_0$ ~  & ~$dh$
 & ~ ~ $\chi^2$ & dof & cl($\%$)\end{tabular}&
\begin{tabular}{llllll}$\Omega_0$ ~ &  $\Omega_{\Lambda0}$~ &
~$dh$ ~ & ~$\chi^2$ & dof & cl($\%$)\end{tabular}\\
\hline

$\Lambda \sim S^{-2}$ & \begin{tabular}{lllll}0.68  & 18.25  &  4.71
& 10 & ~ ~9\end{tabular}&
\begin{tabular}{llllll}0.97 & 0.61 & 17.99 & 4.66 & 9 & ~ ~14\end{tabular}\\

$\Lambda \sim H^2$ & \begin{tabular}{lllll}0.67  & 18.53  &  4.72
& 10 & ~ ~9\end{tabular}&
\begin{tabular}{llllll}0.29 & 1.03 & 23.71 & 4.28 & 9 & ~ ~11\end{tabular}\\

$\Lambda \sim \rho$& \begin{tabular}{lllll}0.67  & 18.53  &  4.72
& 10 & ~ ~9\end{tabular}&
\begin{tabular}{llllll}0.53 & 0.82 & 22.41 & 4.27 & 9 & ~ ~11\end{tabular}\\

$\Lambda=\mbox{const}$& \begin{tabular}{lllll}0.2 ~ & 22.29  &  4.52
& 10 & ~ ~8\end{tabular}&
\begin{tabular}{llllll}0.08 & 1.16 & 27.73 & 4.17 & 9 & ~ ~10\end{tabular}\\

\begin{tabular}{l}Gurvits's\\model \end{tabular}
 & \begin{tabular}{lllll}  &  & & &
\end{tabular}&
\begin{tabular}{llllll}0.43 & 0 ~ & ~17.72 & 4.74 & 10 & ~ ~9\end{tabular}\\

\hline
\end{tabular}
\end{center}


\vspace{.5cm}
We thus find that the Gurvits et al' model (Friedmann model with a vanishing
$\Lambda$) is not the best-fitting model. The best-fitting model with
constant $\Lambda$ is a low-density model dominated by positive
$\Lambda$. It is clear from Table 2 that all the models, variable- or
constant-$\Lambda$, show an excellent fit to the data with
$\chi^2/$dof $\approx 0.5$ and require
non-zero, positive values of $\Lambda$. Here again the
best-fit values of $\Omega_0$ for variable $\Lambda$-models are larger
than the best-fit $\Omega_0$ for the constant $\Lambda$-model.
Unlike the supernovae data, this data set requires both, the accelerating
and decelerating models: the global best-fit parameters (for the
variable as well as the constant $\Lambda$-models) represent an
accelerating expansion whereas the best-fit parameters for the flat models
(only the variable $\Lambda$-ones) represent a decelerating expansion.

Once again we encounter with the same
values of the parameters for the models $\Lambda \sim H^2$ and
$\Lambda \sim \rho$ for the flat cases, which serves a cross check
for the fitting procedure. In Figure 2, we have shown the fit of the
data to the flat models.

\centerline{{\epsfxsize=14cm {\epsfbox[50 250 550 550]{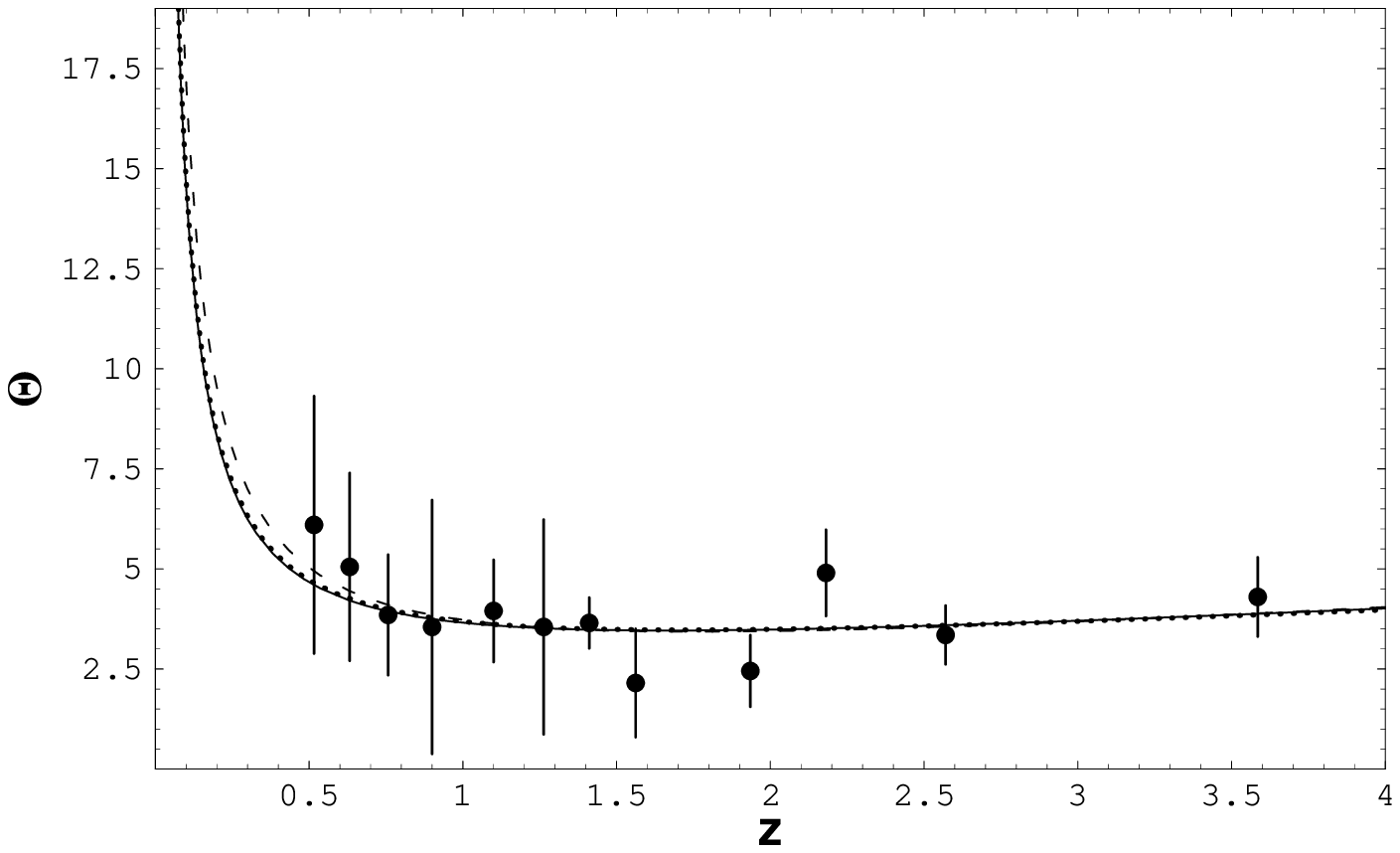}}}}

\noindent
{\bf Figure 2.}
Gurvits et al' data points are compared with the three best-fitting flat
models obtained for the cases: $\Lambda \sim S^{-2}$ with $\Omega_0=0.68$
(solid curve), $\Lambda \sim H^2$ with $\Omega_0=0.67$ (dotted curve) and
$\Lambda=\mbox{constant}$ with $\Omega_0=0.2$ (dashed curve).

\vspace{.8cm}
Following Gurvits et al, it would be worth while to study the consequences
if the source linear size depends on the source luminosity and redshift
(i.e., the sources are not the `true' standard rods). For this purpose, we
consider the following phenomenological expression for $\Theta$ as has been
considered by Gurvits et al:

\begin{equation}
\Theta\equiv d/d_{{\rm A}} \sim \ell h ~ L^\beta (1+z)^\gamma
{\cal D}_{{\rm A}}^{-1},
\end{equation}
where ${\cal D}_{{\rm A}}\equiv H_0 d_{{\rm A}}$ is the dimensionless
{\it angular diameter distance} and $\ell h$ is the linear size scaling factor
which reduces to $dh$ for the no-evolution case. The parameters $\beta$ and
$\gamma$, representing the dependence of the linear size on the source
luminosity and redshift respectively, are to be estimated from the data.
We fit the Gurvits et al' sub-sample to equation (47) for five parameters
$\Omega_0$, $\Omega_{\Lambda0}$, $\ell h$, $\beta$ and $\gamma$. The best-fit
values for the three models are obtained as follows.

\vspace{.4cm}
\noindent
$\Lambda \sim S^{-2}$:

\noindent
$\Omega_0=0.89$, ~ $\Omega_{\Lambda0}=0.58$, ~ $\ell h=23.69$,
~ $\beta=0.37$, ~ $\gamma=-0.45$, ~ $\chi^2=3.21$;

\vspace{.4cm}
\noindent
$\Lambda \sim H^2$:

\noindent
$\Omega_0=0.39$, ~ $\Omega_{\Lambda0}=0.95$, ~ $\ell h=28.16$,
~ $\beta=0.37$, ~ $\gamma=-0.44$, ~ $\chi^2=2.72$;

\vspace{.4cm}
\noindent
$\Lambda \sim \rho$ :

\noindent
$\Omega_0=0.51$, ~ $\Omega_{\Lambda0}=0.83$, ~ $\ell h=17.9$, ~ ~
$\beta=0.38$, ~ $\gamma=-0.42$, ~ $\chi^2=2.47$.

 \vspace{.4cm}
We find that the estimates of $\Omega_0$ and $\Omega_{\Lambda0}$ thus
obtained are in agreement with those shown in Table 2. The estimates
of $\beta$ and $\gamma$ are in qualitative agreement with
$\beta=0.37$ and $\gamma=-0.58$ obtained by Gurvits et al. The values
of $\Omega_0$ and $\Omega_{\Lambda0}$ for the case $\Lambda \sim \rho$
are also in qualitative agreement with the estimates of these parameters
obtained (by assuming no-evolution) from an independent sample of sources
(Vishwakarma 2000b). It thus seems that there is no need to consider
evolutionary effects due to linear size-luminosity or linear
size-redshift dependence. However, this result, which is based on
the above-made comparisons, is
rather superficial than indicative because there are large uncertainties
in the above-mentioned estimates which are due to the large number of
regression parameters.

\vspace{1cm}
\noindent
{\bf 5. Conclusion}

\noindent
We find that all the variable $\Lambda$-models,
fit the data sets equally well and a discrimination between the different
dynamical laws or between the different geometries is not possible with
the present status of the data sets.
While the fit of the models (constant, as well as, variable $\Lambda$)
to the supernovae data is very good with $\chi^2/$dof $\approx 1$, their
fit to the radio sources data is excellent with  $\chi^2/$dof even less than
1.
The estimates of the density parameter $\Omega_0$
(from both data sets) for the variable $\Lambda$-models are found higher
than those for the constant $\Lambda$-Friedmann model. The supernovae data
set requires
an accelerating expansion for the variable $\Lambda$-models as it does
for the constant $\Lambda$-model. However, the radio sources data set
requires accelerating as well as decelerating expansion for the
variable $\Lambda$-models, though only an accelerating expansion for
the constant $\Lambda$-model.

Many of the $\Omega_0$ values shown in Tables 1 and 2 (especially the
global best-fit estimates from the supernovae data) are found rather high.
In view of the small observed value of $\Omega_0$, this implies that
not all the best-fit solutions found here are realistic.
For $0.2 <\sim \Omega_0 <\sim 0.5$, which seems to be the general
consensus at present, the only realistic solutions are (a) estimated from the
supernovae data: the best-fit solutions for flat models (including the
constant $\Lambda$-case); (b) estimated from the radio
sources data: the global best-fit solutions for the models
$\Lambda \sim H^2$ and $\Lambda \sim \rho$, the best-fit solution for
the flat model with $\Lambda=$ constant and the Gurvits et al' model.

As in the case of the recent analyses of the CMB results from BOOMERANG
 and MAXIMA (Jaffe et al, 2000), our global best fit values from both
data sets imply a spherical universe ($\Omega_0+\Omega_{\Lambda0}>1$).

One may notice from Tables 1 and 2 that the best-fit values of $\Omega_0$
and $\Omega_{\Lambda0}$ from the two data sets are a bit too different for
the corresponding models, especially in the non-flat cases. The reason
for this may be the incomplete understanding of a number of astrophysical
effects and processes (evolution, inter galactic dust, etc.).
It should also be noted that the Gurvits et al' compilation is based on
very inhomogeneous data obtained by many different observers using
different instruments and imaging techniques. New VLBI observations now
in progress will improve the accuracy of the observed $\Theta-z$ relations
as it will provide a uniform data set for analysis.
A reasonable hope is that, as the data accumulates and its understanding
improves, such tests coupled with plausible evolutionary hypothesis will
lead to similar results. These results should be consistent with
the CMB measurements and also with the spatial scale of the peak in the
power spectrum of galaxy perturbations which provides a standard ruler
in co-moving space for measuring cosmological parameters
(Broadhurst and Jaffe 2000, Roukema and Mamon 2000).

\vspace{.3cm}

\noindent
{\bf Acknowledgments}

\noindent
The author thanks Professors J V Narlikar and R S Ellis for fruitful
discussions and the Department of Atomic energy, India for support
available in conjunction with the Homi Bhabha Professorship of Professor
J V Narlikar.
Thanks are also due to Dr. L I Gurvits for sending his compilation.

\vspace{.3cm}
\noindent
{\bf References:}

\noindent
Abdel-Rahaman A-M M 1992 Phys. Rev. D {\bf 45} 3492\\
Abdussattar and Vishwakarma R G 1996 Pramana - J. Phys {\bf 47} 41\\
\rule{1.2cm}{0.1mm} 1997 Class. Quantum Grav. {\bf 14} 945\\
Arbab A I and Abdel-Rahaman A-M M 1994 Phys. Rev. D {\bf 50} 7725\\
Arbab A I 1997 Gen. Relativ. Grav. {\bf 29} 61\\
Beesham A 1994 Gen. Relativ. Grav. {\bf 26} 159\\
Berman M S 1991 Phys. Rev. D {\bf 43} 1075\\
Berman M S and Som M M 1990 Int. J. Theor. Phys. {\bf 29} 1411\\
Bertolami O 1986 Nuovo Cimento B {\bf 93} 36\\
Broadhrust T and Jaffe A H 1999 Preprint (astro-ph/9904348), (in press

\hspace{.5cm} with ApJL)

\noindent
Carroll S M Press W H and Turner E L 1992 Ann. Rev. Astron.

\hspace{.5cm} Astrophys. {\bf 30} 499

\noindent
Carvalho J C Lima J A S and Waga I 1992 Phys. Rev. D {\bf 46} 2404\\
Chen W and Wu Y S 1990 Phys. Rev. D {\bf 41} 695\\
Efstathiou G Bridle S L Lasenby A N Hobson M P and Ellis R S 1999

\hspace{.5cm} MNRAS {\bf 303} L47

\noindent
Efstathiou G 1999 Preprint (astro-ph/9904356)\\
Endo M and Fukui T 1977 Gen. Relativ. Grav. {\bf 8} 833\\
Garnavich et al 1998 Astrophys. J. {\bf 509} 74\\
Gurvits L I 1994 Astrophys. J. {\bf 425} 442\\
Gurvits L I Kellermann K I and Frey S 1999 Astron. Astrophys. {\bf 342} 378\\
Jackson J C and Dodgson M 1996 MNRAS {\bf 278} 603\\
\rule{1.2cm}{0.1mm} 1997 MNRAS {\bf 285} 806\\
Jaffe et al 2000 Preprint (astro-ph/0007333)\\
Kellermann K I 1993 Nat. {\bf 361} 134\\
Lau Y K 1985 Aust. J. Phys. {\bf 38} 547\\
Lima J A S and Carvalho J C 1994 Gen. Relativ. Grav. {\bf 26} 909\\
Lopez J L and Nanopoulos D V 1996 Mod. Phys. Lett. A {\bf 11} 1\\
Overduin J M and Cooperstock F I 1998 Phys. Rev. D {\bf 58} 043506\\
Ozer M and Taha M O 1987 Nucl. Phys. B {\bf 287} 776\\
Perlmutter S et al 1999 Astrophys. J. {\bf 517} 565\\
Perlmutter S, Turner M S and White M 1999 Preprint (astro-ph/9901052)\\
Podariu S and Ratra B 1999 Preprint (astro-ph/9910527)\\
Riess A G et al 1998 Astron. J. {\bf 116} 1009\\
Roukema B F and Mamon G A 2000 Astron. Astrophys. {\bf 358} 395\\
Sahni V and Starobinsky A 2000 Int. J. Mod. Phys. D {\bf 9} 373\\
Salim L M and Waga I 1993 Class. Quantum Grav. {\bf 10} 1767\\
Turner M S and White M 1997 Phys. Rev. D {\bf 56} 4439\\
Vilenkin A 1985 Phys. Rep. {\bf 121} 265\\
Vishwakarma R G 1999 Preprint (gr-qc/9912106)\\
\rule{1.2cm}{0.1mm} 2000a submitted\\
\rule{1.2cm}{0.1mm} 2000b Class. Quantum Grav. {\bf 17} 3833 (gr-qc/9912105)\\
Waga I 1993 Astrophys. J. {\bf 414} 436\\
Waga I and Miceli 1998 Preprint (astro-ph/9811460)\\
Weinberg S 1972 Gravitation and Cosmology, John Wiley\\
\rule{1.2cm}{0.1mm} 1989 Rev. Mod. Phys. {\bf 61} 1\\
Wetterich C 1995 Astron. Astrophys. {\bf 301} 321\\
Zlatev I, Waga I and Steinhardt P J 1999 Phys. Rev. Lett. {\bf 82} 896\\

\end{document}